\documentclass[onecolumn,authoryear]{els-mrw} 

\usepackage{amsmath,amssymb,amsfonts,amsthm,makeidx,graphicx}
\usepackage{txfonts}
\usepackage{helvet}

\newcommand{\aap}{Astron.\ Astrophys.\ }
\newcommand{\apj}{Astrophys.\ J.\  }
\newcommand{\apjl}{Astrophys.\ J.\ Lett.\  }
\newcommand{\apjs}{Astrophys.\ J.\ Suppl.\  }
\newcommand{\mnras}{Mon.\ Not.\ R.\ Astron.\ Soc.\ }
 
\newcommand{\nat}{Nature}

\newcommand{\pasj}{Publ.\ Astron.\ Soc.\ Japan\ }
\newcommand{\prd}{Phys.\ Rev.\ D\ }

\newcommand{\jcap}{Journal of Cosmology and Astro-Particle Physics}

\newcommand{\aj}{Astrophys.\ J.\  }
\newcommand{\araa}{Annual Review of Astronomy \& Astrophysics}

\newcommand{\ssr}{Space Science Reviews}

\newcommand{\Hunit}{km/s/Mpc}

\begin{document}

\chapter{Measuring the expansion history of the Universe with cosmic chronometers}\label{chap1}

\author[1,2]{Michele Moresco}%

\address[1]{\orgname{Universit\`{a} di Bologna}, \orgdiv{Dipartimento di Fisica e Astronomia ``Augusto Righi''}, \orgaddress{via Piero Gobetti 93/2, I-40129 Bologna, Italy}}
\address[2]{\orgname{INAF - Osservatorio di Astrofisica e Scienza dello Spazio di Bologna}, \orgaddress{via Piero Gobetti 93/3, I-40129 Bologna, Italy}}

\articletag{Chapter Article tagline: update of previous edition, reprint..}

\maketitle

\begin{glossary}[Glossary]
\term{Cosmic chronometers} Very massive ($\log_{10}(M/M_{\odot})>$10.5--11) and passively evolving galaxies, used to trace the differential age evolution of the Universe.

\term{D4000} Also known as 4000~\AA~ break, it is a discontinuity in the spectrum of galaxies (in particular of passive ones) appearing at 4000 ~\AA~ restframe wavelength due to the blending of several metallic lines; the older and more metallic the population is, the stronger the break will appear in the spectrum.

\term{Full Spectral Fitting} Technique used in astrophysics to analyze the observed spectra of galaxies (or other astrophysical objects) to derive their physical properties, such as stellar age, metallicity, and star formation history. It involves fitting the entire available spectroscopic and photometric data with a model spectrum generated using stellar population synthesis models.

\term{Lick Indices} Set of spectral absorption features used to study the stellar populations and chemical properties of galaxies (and other astrophysical objects). They have been introduced to measure the strength of absorption features associated with elements like hydrogen, iron, magnesium, calcium, and they have been originally defined considering a standard resolution and wavelength range.

\term{Progenitor bias} Observational systematic effect arising when comparing early-type galaxy (ETG) populations at different redshifts, due to the exclusion of the progenitors of younger ETGs observed at low redshift from the high-redshift sample.

\term{Rejuvenation} Process where a galaxy, after a period of low or no star formation, experiences a rise in star formation rate, bringing it back near or within the star-forming main sequence.

\term{Spectral Energy Distribution} Visual representation of the electromagnetic radiation of a source as a function of wavelength. It can include both the magnitudes and the spectrum of the object considered.

\term{Velocity dispersion} Statistical dispersion of the velocity of stars inside a galaxy, which affects the width of absorption lines in its spectrum. It is typically used to characterize physical properties of galaxies, such as their mass.

\end{glossary}

\begin{glossary}[Nomenclature]
\begin{tabular}{@{}lp{34pc}@{}}
CMB &Cosmic Microwave Background\\
SNe &Type Ia Supernovae\\
BAO &Baryon Acoustic Ocillations\\
CC &Cosmic Chronometers\\
ETG &Early-Type Galaxies\\
IMF &Initial Mass Function\\
SED &Spectral Energy Distribution\\
SPS &Stellar Population Synthesis\\
SFH &Star Formation History\\
SNR &Signal to Noise Ratio\\
sSFR &specific Star Formation Rate\\
\end{tabular}
\end{glossary}

\begin{abstract}[Abstract]
As revealed by Hubble in 1928, our Universe is expanding. This discovery was fundamental to widening our horizons and our conception of space, and since then determining the rate at which our Universe is expanding has become one of the crucial measurements in cosmology. At the beginning of this century, these measurements revealed the unexpected behavior that this expansion is accelerating and allowed us to have a first glimpse of the dark components that constitute $\sim$95\% of our Universe. Cosmic chronometers represent a novel technique to obtain a cosmology-independent determination of the expansion of the Universe, based on the differential age dating of a population of very massive and passively evolving galaxies. Currently, with this new cosmological probe it is possible to constrain the Hubble parameter with an accuracy of around 5\% at $z\sim0.5$ up to 10-20\% at $z\sim2$. In this Chapter, the cosmic chronometers approach is presented, describing the method and how an optimal sample can be selected; it is then discussed how the most recent measurements of the expansion history of the Universe have been obtained with this approach, as well as the cosmological constraints that can be derived. Particular attention will be given to the systematics involved in this approach and the treatment to properly take them into account. We conclude by presenting forecasts that show how future spectroscopic surveys will significantly boost the accuracy of this method and open the possibility to a percent determination of the Hubble constant, making cosmic chronometers a powerful independent tool to derive information on the expansion history of the Universe.

\end{abstract}

\section{Introduction}\label{chap1:sec1}

Until a century ago, our Universe was a relatively stable, calm, and well-understood place to live in; or at least so we thought. In this scenario, the observational results by \cite{hubble1929}, previously anticipated by the theoretical works of \cite{lemaitre1927} and \cite{robertson1929}, have been one of the breakthroughs of modern astrophysics, opening a new and unexpected door. Instead of being static, the Universe was expanding, with speed $v$ (at distances large enough so that proper motions are negligible) proportional to the distance $d$ of the tracer considered:
\begin{align}\label{chap1:eq1}
v=H_0\cdot d\;\;
\end{align}
Hubble's analysis also provided the first measurement of this expansion rate by measuring the proportionality constant of this relation, which was afterward named after the same astronomer the ``Hubble constant''. Since then, measuring not only locally, but also at larger and larger distances the rate at which the Universe is expanding has been one of the main goals of cosmology because it can provide crucial information on the components of our Universe and on the balance between the forces in place.

Another shock to the pillars of modern cosmology was given around the beginning of this century, when two independent groups measuring the distances of distant supernovae determined that not only is the Universe expanding, but also that this expansion is accelerating \citep{riess1998, perlmutter1998, perlmutter1999}. This discovery unveiled the presence of an unknown form of energy, named ``dark energy'', responsible for this accelerated expansion, and, starting from the early 2000s, several techniques, methods, and surveys have been developed to understand its nature. Currently, it is still unclear whether this acceleration is due to an unknown form of energy or to a modification of General Relativity on very large scales, and different models have been explored and proposed to address it. To better understand it, several cosmological probes have been introduced, studied, and improved to constrain the properties of the components of the Universe with increasing accuracy. Many of those, supported by technological advances and dedicated surveys, have now become standard in the framework of modern cosmology, such as the analysis of the Cosmic Microwave Background (CMB), Type Ia Supernovae (SNe), and Baryon Acoustic Oscillations (BAO); the reader can find an extensive review on these topics in \cite{huterer2018} (and references therein).

With time, technological, and observational improvements, several cosmological probes reached the golden goal of percentage and sub-percentage accuracy on several cosmological parameters, e.g., the Hubble constant $H_0$. This gave a revamped attention to the determination of the local value of the expansion rate of the Universe, since, differently from the expectations, measurements obtained with different methods did not converge to a common value, but presented an increasing discrepancy, which is now at the level of 4-5$\sigma$. This is currently known as the ``Hubble tension'' \citep{verde2019}. Whether this tension is driven by some unaccounted systematic effect or it is giving hints that new physics is required is still under debate \citep[for an in-depth review on the topic, see][]{abdalla2022}. 

It is clear that to make steps forward it is crucial to go beyond the standard cosmological probes presented above. Having multiple, independent, and complementary probes will allow us on the one side to keep systematics under control, and on the other side to increase the accuracy of the derived cosmological parameters, taking advantage of the strength of each probe. Several emerging cosmological probes have been proposed during the last years, including gravitational waves as standard sirens, quasars and gamma-ray bursts as standard candles, time-delay cosmography, surface brightness fluctuations, and redshift drift. Many of those are now reaching very promising results, and the interest in those of the scientific community is increasing with the advent of new telescopes and gravitational waves observatories \citep[see][ for a review on these emerging cosmological probes]{moresco2022}. The synergy between these new methods will become crucial for the future of modern cosmology, opening new avenues and providing pieces of evidence fundamental to tackling the various tensions \citep[see Sect. 4 of][]{moresco2022}. Here, we will discuss a novel cosmological probe named ``cosmic chronometers'' (CC), that can provide us a new way to derive a cosmology-independent measurement of the expansion history of the Universe.

\section{What are cosmic chronometers?}\label{chap1:sec2}
A chronometer is an instrument to measure time. Differently from a clock, it is optimized to provide very accurate differential timing measurements, independently of absolute time. The cosmic chronometer method relies on the exact same principle.\\ 
The expansion rate of the Universe, also known as Hubble parameter $H(z)$, is defined as:
\begin{equation}
H(z)=\frac{\dot{a}}{a}\;\;,
\label{eq:Hz_a}
\end{equation}
where $a(t)$ is the adimensional scale factor, relating physical $R(t)$ and comoving $r$ distances, $R(t)=a(t)r$.\\ 
The idea behind the CC method, originally proposed by \cite{jimenez2002}, is that the measurement of the differential age evolution of the Universe $dt$ in a redshift interval $dz$ (i.e., how much the Universe has aged between two redshift bins) could be used to provide a direct determination of $H(z)$. With only the minimal assumptions of a Friedmann-Lema{\^\i}tre-Robertson-Walker (FLRW) metric, the scale factor can be written as $a=1/(1+z)$, and hence combining this equation with Eq.~\ref{eq:Hz_a} it can be obtained that:
\begin{align}
H(z)=-\frac{1}{(1+z)}\frac{dz}{dt}\;\;.
\label{eq:Hz}
\end{align}
It is important to underline here two main points: (i) in this equation, apart from the FLRW metric, no further cosmological assumption is considered, so this technique provides a cosmology-independent estimate of the Hubble parameter; (ii) as can be inferred from Eq.~\ref{eq:Hz}, in this method the only relevant quantity will be the differential time evolution $dt$, and not the absolute age (from which the name ``cosmic chronometers'' arises). The fundamental step will be to find a way to constrain the differential age evolution of the Universe robustly. Different approaches can be exploited for this purpose.

A first solution could be to derive stellar ages for a wide population of objects at various redshifts, consider their distribution, and determine, at each redshift, the objects older than some percentile threshold (to avoid potential biases by outliers); the obtained upper, oldest envelope\footnote{This method in the literature is also referred to as red envelope since the oldest galaxies are also the one with redder colors.} of galaxies can then be used to measure the differential $dt/dz$. This approach, while having the advantage of relying on an easier sample selection (i.e., all galaxies observed, since the younger ones would not affect the position of the upper envelope), requires, however, a very high number of objects to robustly recover the upper envelope, and a very complete sample not to bias the slope of the age-redshift relation \citep[see, e.g., ][]{jimenez2002, jimenez2003,simon2004,moresco2012}. It is, therefore, not convenient observationally speaking.

The alternative approach is, instead, to base the analysis on objects able to trace accurately the ageing of the Universe as a function of redshift, i.e. cosmic chronometers. The ideal goal is to find a population of astrophysical sources with a homogenous and synchronized formation time, that assembled most of their mass at very high redshifts and that evolved without major further episodes of star formation since then so that they could be used to accurately map the ticking of cosmic time.
Since the discovery of bimodality in galaxy morphologies between elliptical and spiral \citep[the Hubble's tuning fork,][]{hubble1936}, it was found that the observed sequence corresponded also to a difference in star formation rates and history \citep{roberts1963}. In the last decades, a wide literature converged to the picture that very massive ($\log_{10}(M/M_{\odot})>$10.5--11), elliptical, passively evolving galaxies could represent the best CC because of their very homogeneous properties.
These objects have been found to have formed their mass in very rapid episodes of star formation \citep[on a timescale of the order of $\Delta t<0.3$ Gyr,][]{thomas2010,mcdermid2015,citro2017,carnall2018} and at high redshifts \citep[$z>2-3$,][]{franx2003,cimatti2004, daddi2005,choi2014,mcdermid2015,onodera2015,pacifici2016,carnall2018,estrada2019,carnall2019,kriek2019,belli2019}; their generally accepted formation scenario is that they have consumed all their gas reservoir very quickly, and that they are evolving passively through cosmic time since this burst of star formation. To support this picture, they have been found to have also a very homogeneous stellar metallicity from solar to slightly oversolar from the local Universe up to $z\sim2$ \citep{gallazzi2005,onodera2012,gallazzi2014, conroy2014,onodera2015,mcdermid2015,citro2016,comparat2017,saracco2019,morishita2019,estrada2019,kriek2019}. This population of massive and passive galaxies has been found extensively up to high redshifts ($z \gtrsim 2.5$) \citep[][]{daddi2004,fontana2006,ilbert2006,wiklind2008,caputi2012,castro2012,muzzin2013,stefanon2013,nayyeri2014,straatman2014,wang2016,mawatari2016,deshmukh2018,merlin2018,merlin2019,girelli19}, but there is also evidence of their presence up to $z>7-9$ \citep[$\sim$500 Myr after the Big Bang, e.g., ][]{labbe2023,adams2023}; they represent, for this reason, the oldest galaxy population at each redshift. Moreover, it was also observed that more massive galaxies are formed earlier than less massive ones and over shorter timescales, being therefore very homogeneous in terms of formation redshift; this scenario is often referred to as ``mass downsizing'' \citep{Heavens2004,cimatti2004,thomas2010}.
Having identified a potential population of qualified CCs, the next step is to define how to accurately select them, and how to measure their differential ages.

\begin{figure}[t]
\centering
\includegraphics[width=.95\textwidth]{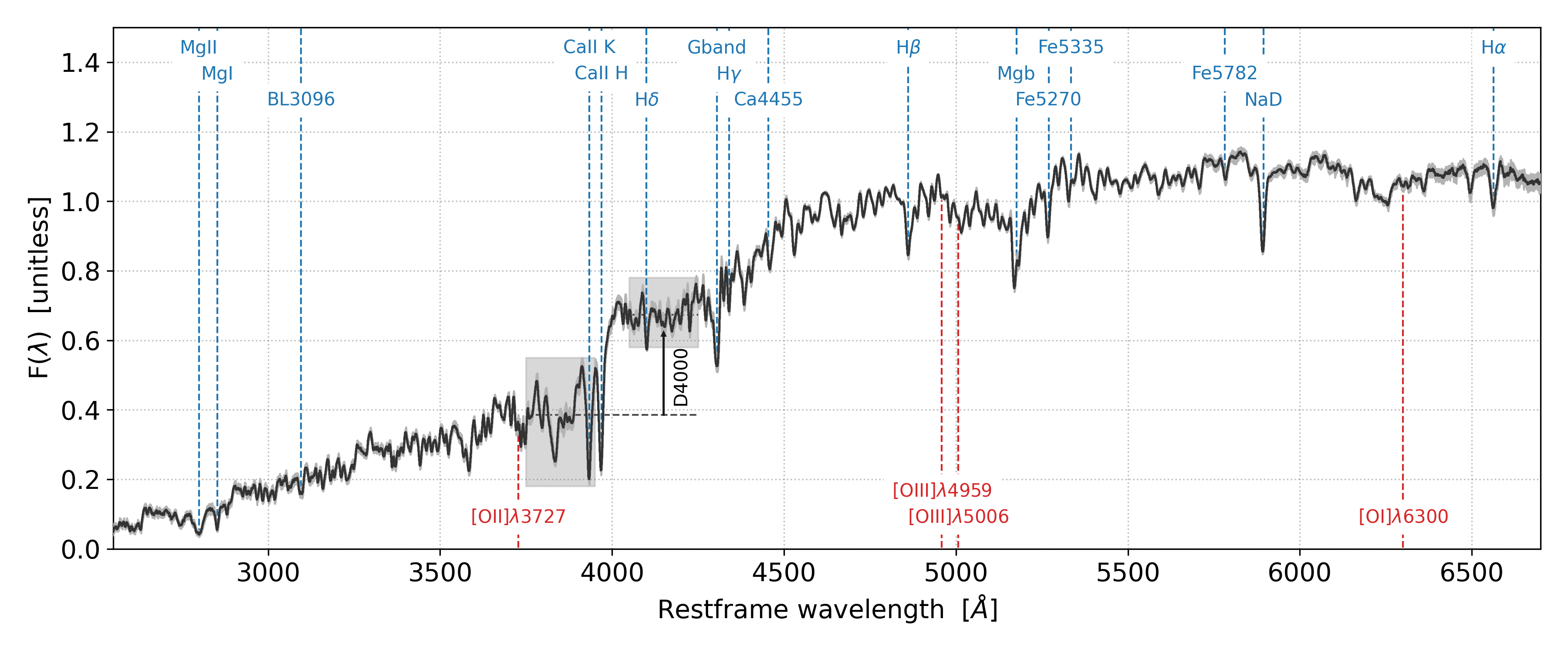}
\caption{Stacked spectrum of a sample of massive and passively evolving galaxies, obtained from the combination of $\sim10^5$ spectra from the Sloan Digital Sky Survey. It is evident the red continuum, as well as the typical absorption features characterizing the spectra of passive galaxies (shown with blue vertical dotted lines), and the total absence of emission lines (their positions are shown by the red dotted lines). In the plot, it is also highlighted another feature distinctive of the spectra of passive galaxies, the 4000 $\AA$ break (D4000); this discontinuity is caused by the accumulation of several metallic lines in old galaxy populations. The stacked spectrum is normalized to unity in the region $4500<\lambda_{\rm rest}\;[\AA]<5000$.}
\label{fig:stackedspectrum}
\end{figure}

\section{How to select cosmic chronometers?}\label{chap1:sec3}

\begin{figure}[]
\centering
\includegraphics[width=.95\textwidth]{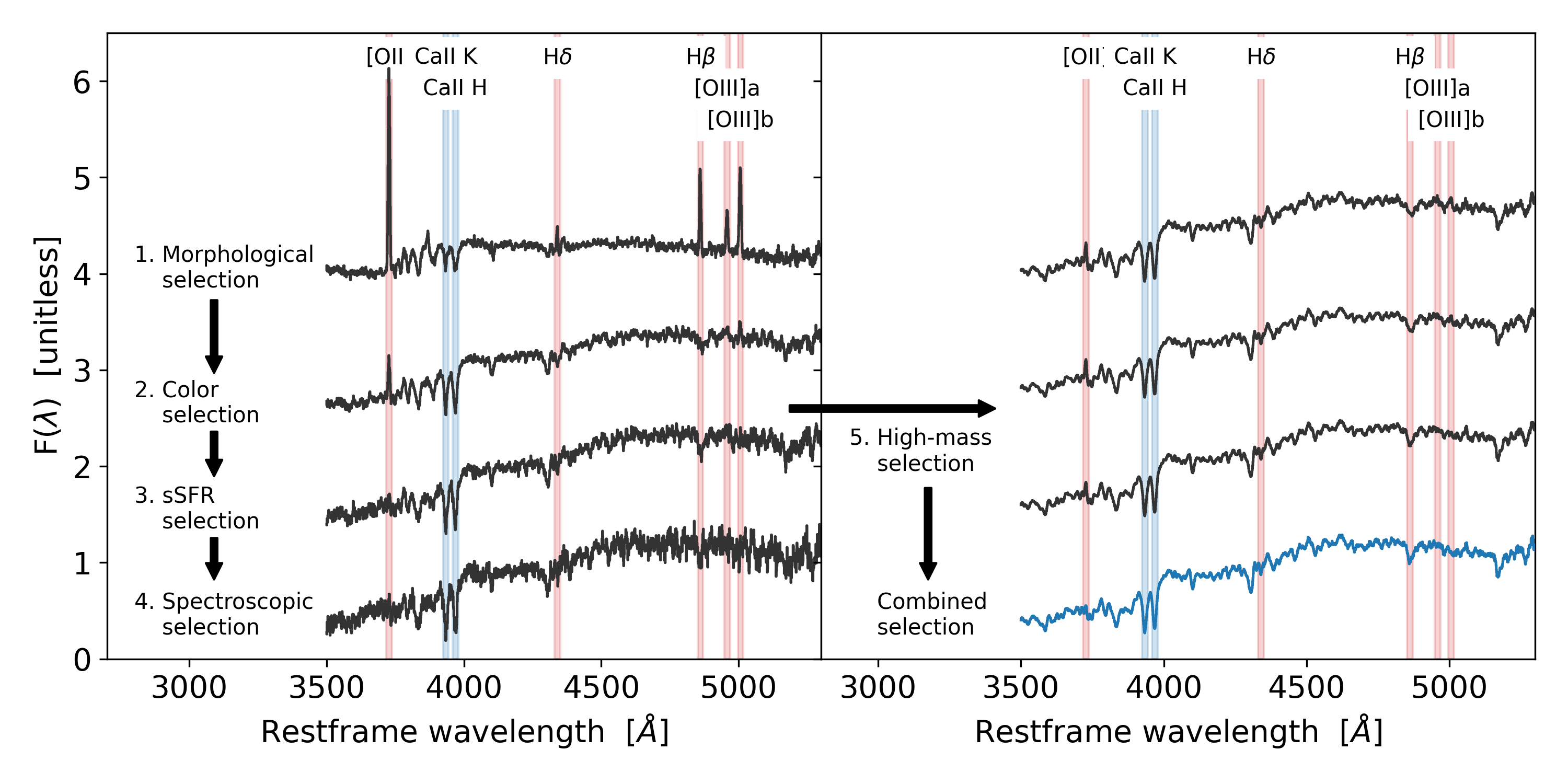}
\caption{Cosmic chronometers selection workflow. In the plots are shown the stacked spectra \citep[from][]{moresco2013} of differently selected samples of CC, as discussed in Sect.~\ref{chap1:sec3}; an arbitrary offset has been applied to each spectrum to improve their visibility. In the left panel are shown the spectra of ``passive'' galaxies selected with the various criteria and in the right panel their analogous ones where also a cut in stellar mass ($\log_{10}(M/M_{\odot})>$10.75) is combined to each selection criterion. Finally, the blue spectrum in the bottom right corner represents the spectrum of the purest CCs, where all the above criteria have been combined. Vertical red and blue shaded regions represent the position of the main emission (tracers of potential ongoing star-formation, hence contamination) and absorption features, respectively, as discussed in the text.}
\label{fig:stacked_selection}
\end{figure}

As introduced in the previous section, the first criterion that we may think about to obtain a sample of cosmic chronometers is to select quiescent galaxies. This population, sometimes inappropriately referred to as early-type galaxies\footnote{There is a wide literature on the topic, and for an extensive review it is possible to see \cite{renzini2006} and references therein.}, can and has been defined with very different selection criteria during the years, based on morphology, or colors, or spectroscopic properties \citep[for a more extensive discussion, see][]{moresco2022}. However, not all criteria are equivalent, and several works have shown that depending on the choice one will end up with a sample with different levels of contamination by star-forming objects, ranging from 10 up to 50\% \citep{franzetti2007,moresco2013,belli2017,schreiber2018,fang2018,merlin2018,leja2019,diaz2019}.

In the CC approach, one of the key points is to obtain the purest possible sample to be able to derive an unbiased differential age estimate, because different levels of contamination at different redshifts may affect the slope of the age-redshift relation, hence the determination of the Hubble parameter. It is therefore pivotal to select cosmic chronometers accurately removing any possible residual source of star-forming outliers. Several works highlighted that the best approach to reach this goal is not to rely on a single criterion, but to combine as much information as possible, including morphological, photometric, spectroscopic, and stellar-mass cuts \citep{moresco2013,borghi2022,tomasetti2023}.

\begin{BoxTypeA}{Cosmic chronometers selection process}

\noindent To ensure the necessary purity in a sample of cosmic chronometers, it is not enough to just select massive and passive galaxies, and not even to base the selection on a single criterion, but different methods should be combined. The fundamental steps, discussed by \cite{moresco2018}, are described below, and their effects are shown in Fig.~\ref{fig:stacked_selection}.\\~\\

\begin{enumerate}
\item {\bf Morphological selection.} Elliptical galaxies have been identified for a long time to be older and more passive compared to spiral ones \cite[see, e.g.,][]{roberts1963}, therefore the first criterion that we could explore is the morphological one. However, the issue with this selection is that it has been found that the morphological timescale (the time needed to turn a galaxy into a red spheroidal) and the quenching timescale (the time needed to shut the star formation history) are different, and delayed up to  1-2 Gyr \citep{pozzetti2010}. For this reason, as found in \cite{franzetti2007} and \cite{moresco2013}, this selection is also the least effective to get an uncontaminated sample, and also the one less affecting the final result if not implemented. The next ones are, on the other hand, fundamental.
\item {\bf Color selection.} As shown in Fig.~\ref{fig:stackedspectrum}, the typical spectrum of a passive galaxy has a red continuum. Different color-color selections have been proposed in the literature, typically based on magnitudes spanning a wide range of wavelengths (from the ultraviolet to the infrared bands, based on the availability) to better capture the slope of the spectrum. The most commonly used are the NUVrJ diagram \citep{ilbert2013}, the UVJ diagram \citep{williams2009}, and the NUVrK \citep{arnouts2013}; other possibilities are based on the full shape of the Spectral Energy Distribution \citep[e.g.,][]{ilbert2009,zucca2009}.
In this case, it is of particular importance to have magnitudes that provide information on the UV emission, to disentangle between objects intrinsically red, or reddened due to extinction, and to detect potential contamination of a young population (from 100 Myr to 1 Gyr); from this point of view, the NUVrJ selection has been identified as one of the best, minimally affected by these issues.
\item {\bf specific Star Formation Rate selection.} Since we want to select galaxies that have ceased their star formation, another indicator that can be explored is the specific Star Formation Rate ($sSFR$), i.e. the rate at which a galaxy is forming stars per unit mass ($SFR/M$). This quantity can be either derived from a fit to the Spectral Energy Distribution (SED) of galaxies or from other indicators. In the literature, typically a threshold $\log_{10}(sSFR)<-2\;[Gyr^{-1}]$ has been proposed to separate star-forming and quiescent galaxies \citep{ilbert2010,pozzetti2010,ilbert2013}. This cut corresponds to select galaxies that are increasing their mass for less than 1/100th of their present mass per Gyr.
\item {\bf Spectroscopic selection.} Emission lines have been historically identified as important tracers of potential ongoing star formation (even if also other processes could trigger the presence of emission lines in red galaxies, like low accretion-rate AGN, fast shocks, low-ionization nuclear emission-line regions, or photo-ionization by old post-asymptotic giant branch, see, e.g., \citealt{yan2006,annibali2010, yan2012}). The most studied emission lines are lines corresponding to the Balmer sequence or to oxygen emission, namely [OII]$\lambda$3727, H$\beta$ ($\lambda=4861$\AA), [OIII]$\lambda$5007, and H$\alpha$ ($\lambda=6563$\AA); the line (or lines) on which to base the selection will depend on the spectroscopic data availability and wavelength coverage (and, indirectly, on the redshift). Different criteria can be implemented, based on the integrated flux of the line, on its equivalent width (EW), or on the measured signal-to-noise (SNR) ratio. Typical cuts adopted are EW$<5\;\AA$ \citep{mignoli2009, moresco2012, borghi2022}, SNR$<$3--5 \citep{moresco2016,wang2018}, or also a combination of these \citep{moresco2016,tomasetti2023}. Whichever selection is chosen, it is crucial, to be conservative, that the final spectra do not show any evidence of emission lines (see, e.g. Fig.~\ref{fig:stacked_selection}). In parallel to emission lines, recently also another spectral indicator has been proposed, based on the ratio between CaII H and K lines \citep{moresco2018}. In particular, the ratio of these two lines can trace the presence of contamination by a young stellar component since the H$\epsilon$ line at 3970 $\AA$ gets combined to the CaII H line, inverting the standard ratio for which the CaII K line is deeper than the CaII H line. \cite{borghi2022} demonstrated that this indicator correlates extremely well with other indicators of ongoing star formation (NUV colors, SFR, presence of emission lines), so an additional selection on this diagnostic could help to improve the quality of the selected sample \citep[as done, e.g., in][]{borghi2022, jiao2023, tomasetti2023}.
\item {\bf High-mass (/velocity dispersion) selection.} As discussed above, many works found that objects with higher masses are also the oldest ones, most coeval, and with the most synchronized SFH \citep[mass-downsizing, see e.g.][]{cimatti2004, thomas2010}. Moreover, in line with these findings, it has also been found that imposing a threshold to select galaxies with large stellar mass (or velocity dispersion $\sigma$, that correlates with the stellar mass in early-type galaxies) significantly improves the purity of the selected CC sample, decreasing the contamination by star-forming outliers \citep[see, e.g.,][]{moresco2013}. Typically, a cut $\log_{10}(M/M_{\odot})>$10.6--11 (or $\sigma>180-200$ km/s) is considered to select CC.
\item {\bf Combined selection.} In the end, as demonstrated by several works \citep{moresco2012, moresco2016, borghi2022, jiao2023, tomasetti2023}, to guarantee the selection of a pure sample of massive and passively evolving galaxies, several of the above criteria have to be combined, depending on the availability of photometric, spectroscopic, and morphological data. As a good rule of thumb, at least the color, spectroscopic, and mass selection have to be combined, since, looking at complementary features, they ensure to minimize the contamination of the sample. A full selection diagram for CC selection is provided in \cite{moresco2018} and \cite{moresco2022}.
\end{enumerate}
\end{BoxTypeA}

Fig.~\ref{fig:stacked_selection} provides an example of the application of the different selection criteria discussed above, showing the stacked spectra of quiescent objects as obtained from different definitions \citep[from][]{moresco2013}. Clearly, a simple color or morphological selection is not enough, since they leave a significant contamination of a younger star-forming population, as can be seen from the clear presence of strong emission lines. On the other hand, a simple sSFR or spectroscopic selection, even if they remove the strongest emission lines, does not provide a completely uncontaminated sample, as can be seen by the CaII H/K lines ratio. At the same time, the panel on the right shows the strong impact on the spectra of including in the selection also a high mass cut ($\log_{10}(M/M_{\odot})>$10.75) to the previous criteria, since all the strong emission lines are significantly reduced. In the end, the combination of photometric, spectroscopic, and mass selection has been proven to be the most effective in obtaining the purest possible sample of CC.
The impact of each criterion in terms of number of objects excluded in the final combination may depend on the selection function of the starting sample \citep[as an example of this difference, the reader can compare][]{moresco2016,borghi2022,tomasetti2023}.

\section{Measuring relative ages}\label{chap1:sec4}
Once an optimal sample of CCs has been selected, it is possible to focus on measuring the relevant quantities needed for the CC method. In Eq.~\ref{eq:Hz} there are two unknowns: the redshift interval $dz$ and the differential age $dt$. With the advent of large spectroscopic surveys, either ground- or space-based\footnote{e.g., the Sloan Digital Sky Survey \citep[SDSS,][]{SDSS}, the Dark Energy Spectroscopic Instrument \citep[DESI, ][]{DESIa, DESIb}, and the ESA space mission Euclid \citep{Euclid}, just to make a few examples.}, determining the redshifts with very high accuracies (of the order of $0.001\times(1+z)$ and better) has become a relatively easy task. Therefore, in the above equation, $dz$ can be accurately determined.

Measuring the ages of a stellar population in astronomy is, on the other hand, a whole other story. The main problem is that there are physical parameters of galaxies that affect the shape of their spectrum in degenerate ways, so that a change of one parameter can be compensated by the variation of the other one to produce a very similar spectrum; among those, the most significant one is the age--metallicity degeneracy \citep{worthey1994,ferreras1999}, because both an older age and a higher metallicity creates a redder spectrum. For this reason, even if
widely used in the literature, stellar ages based on the fit of simply the magnitudes of a galaxy are less accurate and robust, being more prone to this kind of degeneracies, and would require very accurate calibration to be used in this context. On the other hand, the full spectrum of galaxies contains plenty of useful information that can be used to break these degeneracies and obtain accurate relative ages.

\begin{figure}[t]
\centering
\includegraphics[width=.95\textwidth]{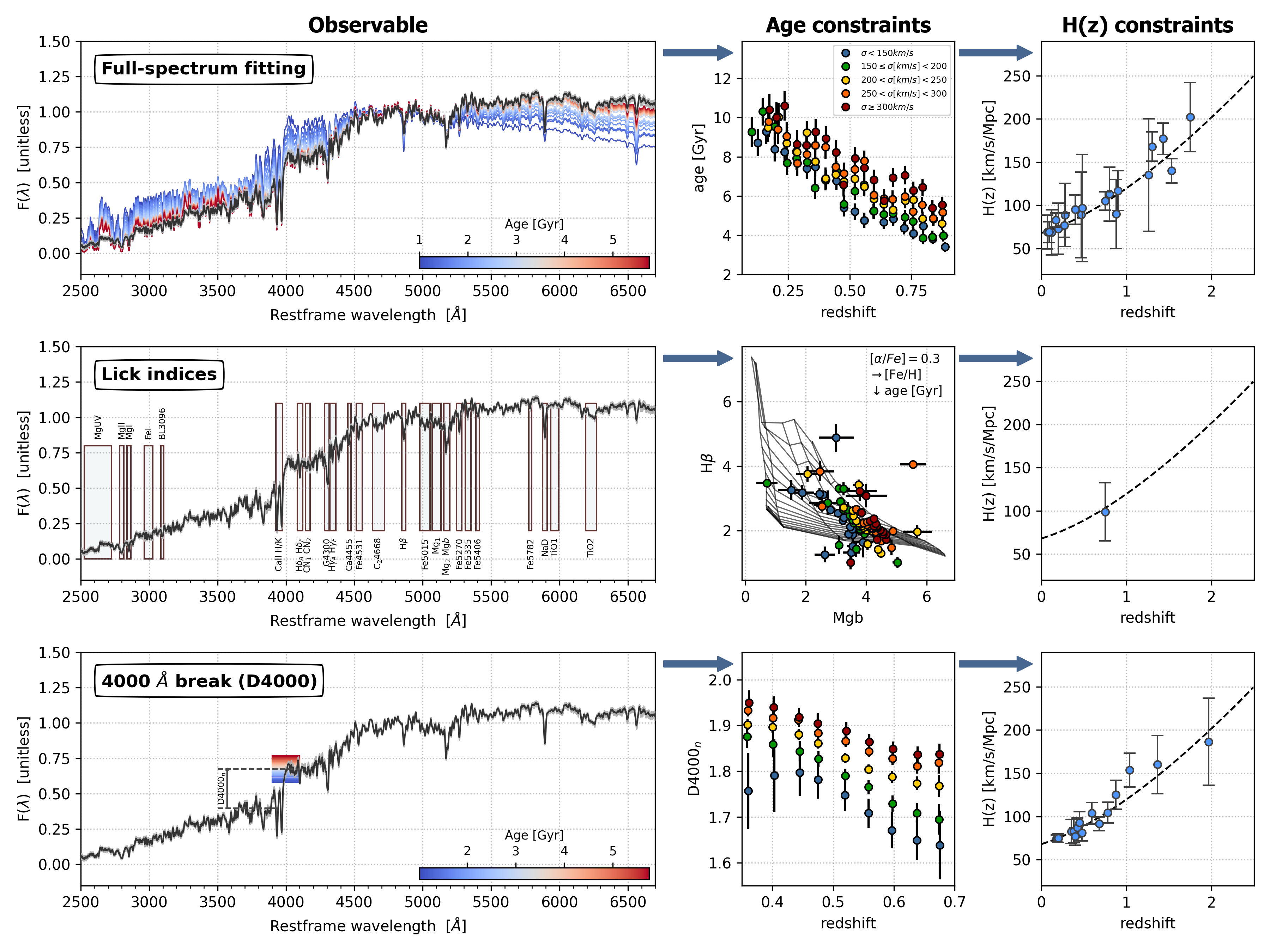}
\caption{Different methods used to derive (relative) ages of CC and corresponding cosmological constraints. The various rows show the various methods discussed in the text, namely the Full-spectrum fitting (first row), Lick indices measurements (second row), and D4000 analysis (third row). The first column presents the observable considered (the complete spectrum for the FSF, the main absorption features for Lick indices, and the 4000 $\AA$ break for the D4000), as well as the theoretical models that are compared with the observable to derive the age constraints. In the second column, the derived age constraints are shown, either in the form of the age--redshift relation (top panel) or of the spectral feature used to trace the age (bottom panel), or as they can be inferred compared to the model grid (in middle panel). The spectrum shown is the same as of Fig.~\ref{fig:stackedspectrum}, while the age constraints are simulated data, provided only for visual representation. The third column represents the real corresponding $H(z)$ constraints that have been retrieved in the literature with each method (see Sect.~\ref{chap1:sec5}). }
\label{fig:CCmethods}
\end{figure}

\begin{BoxTypeA}{Methods to derive the differential age evolution of cosmic chronometers}

\noindent We describe here the three main methods that have been used in the literature to derive stellar ages, and hence relative ages, in the CC approach. These methods are also shown in Fig.~\ref{fig:CCmethods}, which provides a visual representation of all of them.\\~\\

\begin{itemize}
\item {\bf Full spectrum fitting.} The SED of a galaxy contains plenty of information about its physical properties, mainly in the form of its continuum and absorption lines. In general, electromagnetic radiation can be averaged over a window in wavelengths to compute the magnitudes or be represented in a more detailed way through its spectrum. To extract information from the SED of galaxies, Stellar Population Synthesis (SPS) models have been created; these are theoretical models that, combining the physical properties and recipes for the formation and evolution of stars, can reproduce the integrated spectrum, and in general the SED, of the galaxy considered. The Full Spectral Fitting (FSF) method is a statistical method that compares the observed SED of a galaxy with a grid of SPS templates, finding the one that better reproduces the observation, and hence deriving the property of the source (i.e. stellar mass, metallicity, age, SFH, ...); differently from a pure spectral fitting, in this case typically a combination of spectrum and magnitudes are considered, maximizing the information content considered from the source. Taking advantage of the full spectral information, the benefit of this approach is that it can provide a joint constraint on all properties, and the larger the wavelength coverage, SNR, and resolution, the more information that can be retrieved and degeneracies between parameters broken. Several codes are available to perform a FSF, which can differ on the statistical method implemented and/or on the SPS models included; among most used, there are \texttt{MOPED} \citep{MOPEDI,MOPEDII}, \texttt{STARLIGHT} \citep{cidfernandes2005}, \texttt{VESPA} \citep{tojeiro2007}, \texttt{ULySS} \citep{koleva2009}, \texttt{BEAGLE} \citep{chevallard2016}, \texttt{FIREFLY} \citep{wilkinson2017}, \texttt{pPXF} \citep{cappellari2017}, and \texttt{BAGPIPES} \citep{carnall2018}.

\item {\bf Lick indices.} Instead of considering the full spectrum, a different approach proposed in the literature is to take advantage only of the absorption features of the spectrum, since they hold significant information about the properties of the galaxy. As highlighted in the seminal works of \cite{worthey1994} and \cite{worthey1997}, different chemical elements are responsible for different absorption lines, and hence can be used to trace different physical properties. In particular, the Balmer lines produced by the hydrogen are very good tracers of the age of the stellar population, iron lines can be used to retrieve information about the metal content, while other lines (like magnesium-related lines) are useful to constrain the enhancement in alpha-elements. \cite{worthey1994} and \cite{worthey1997} introduced standard methods to measure these lines, named Lick indices, and quantified also their dependence on stellar age and metallicity. An essential step for the quantitative interpretation of these data was made possible by the work of \cite{thomas2011}, followed later by \cite{conroy2012} and \cite{vazdekis2015}, who provided models at high resolution for Lick indices depending on age, metallicity and with variable element abundance ratios; these models allow us to derive age constraints by comparing the measured indices with the provided grid, breaking the age-metallicity degeneracy through the combination of different indices \cite[for a discussion on the impact of different indices to be analyzed on the results, see][]{borghi2022}. It is worth noting that public suites exist also to measure Lick indices, such as \texttt{indexf} \citep{cardiel2010} and \texttt{pyLick} \citep{borghi2022}. 
\item {\bf 4000~\AA~break (D4000).} A different method to derive age constraints for CC has been introduced by \cite{moresco2011}. In this work, it was proposed to simplify the approach, and, instead of studying the entire spectrum or several absorption features, to focus on only one that can trace accurately the aging of the stellar population. In particular, it was suggested to use the break at 4000~\AA~rest-frame ($D4000$, a characteristic feature of the spectra of massive and passive old galaxies, see Fig.~\ref{fig:stackedspectrum}). This spectral feature is generated by the blend of several metallic lines below 4000~\AA~that is generated at increasing age and metallicity of the stellar population. Different definitions have been proposed to measure the $D4000$, considering the ratio of the flux in larger or narrower windows across the break \citep{bruzual1983,balogh1999}. This approach, later refined in \cite{moresco2012}, is based on the fact that this feature was demonstrated to be extremely well-correlated (at fixed metallicity) with the age of the galaxy, with a very simple relation:
\begin{equation}
D4000=A(Z, SFH)\cdot{\rm age}+B\;\;.     
\label{eq:D4000age}
\end{equation}
In this equation, the parameter $A(Z, SFH)$ measures the slope of the $D4000$--age relation (depending on the metallicity of the population and, less significantly, on the SFH), and should be calibrated with SPS models. The parameter $B$ is, instead, of no interest, because, as stressed before, the CC method relies on differential measurements. In particular, if we differentiate Eq.~\ref{eq:D4000age}, we obtain a linear relation between the relative evolution in $D4000$ and the relative change in age of the population, $dD4000=A(Z, SFH)\times{\rm dt}$. When we combine this equation with Eq.~\ref{eq:Hz}, we get:
\begin{equation}
H(z)=-\frac{A(Z, SFH)}{1+z}\frac{dz}{dD4000}\;\;.
\label{eq:CC2}
\end{equation}
Compared to Eq.~\ref{eq:Hz}, this new equation has the advantage that statistical and systematic errors have been decoupled, since $dz/dD4000$ is a purely observational quantity to be estimated from the data and $A(Z, SFH)$ is a parameter that encompasses all systematic effects and model dependencies. This method, successfully applied in several works (see Sect.~\ref{chap1:sec5}), is described in detail in \cite{moresco2022}.
\item {\bf Machine-Learning based approach.} In recent years, another approach has been explored to derive stellar ages based on Machine Learning \citep{jimenez2023}; in this work, a neural network has been trained on precise age measurements obtained from Lick indices \citep{borghi2022} to infer ages based only on photometric data. With this approach, which significantly improves over a simple SED-fitting technique, it was possible to estimate the ages of a wide sample of $\sim20000$ passively evolving galaxies and derive accurately the upper envelope of the obtained age--redshift distribution. This analysis opens a new possible way to complement the previous ones, in particular promising since in the near future Euclid \citep{Euclid} and the Vera Rubin Observatory \citep{LSST} will provide a large set of photometric data for billions of galaxies.
\end{itemize}
\end{BoxTypeA}

A visual representation of each method is given in Fig.~\ref{fig:CCmethods}, where, for each of the three discussed approaches to determine ages, we show the observable and the model used to derive the age measurements, the obtained age-redshift relations, and the derived $H(z)$ constraint (from left to right). The first row presents the results that can be obtained from the FSF, where the stacked spectrum of a CC sample is shown superimposed with different SPS models with varying ages, to illustrate how the method works. The ages, obtained from single or stacked spectra of a CC sample, are then used to build an age--redshift relation (second column), here divided into different velocity dispersion bins. From the age--redshift relation in each mass (/velocity dispersion) bin, the quantity $dz/dt$ is derived (as discussed in Sect.~\ref{chap1:sec5}), from which $H(z)$ is estimated. In the spectrum of the second row is highlighted the position of all the main Lick indices, and in the central panel it is shown how a combination of different indices, compared with the grid of theoretical models \citep{thomas2011}, can yield a measurement of age (changing between horizontal lines in the grid) and metallicity (vertical lines). The last row focuses on the $D4000$, showing how, at fixed metallicity, the amplitude of the $D4000$ can be related to the age of the population. The central panel at the bottom presents the $D4000$--redshift relation, from which (similarly to the FSF case), the quantity $dz/dD4000$ is calculated and, through Eq.~\ref{eq:CC2}, the Hubble parameter is derived.

\begin{BoxTypeA}{A few important remarks on the CC method}

\noindent It is important to stress a few important remarks, that apply to all methods to derive relative ages just presented.

\begin{remark}
As underlined while discussing Eq.~\ref{eq:Hz}, the CC technique is based on a {\bf differential approach} where the relevant quantities are {\bf relative ages}, and not absolute ones. An important consequence of being independent of absolute ages is that any potential systematic bias connected to the estimate of the absolute ages will be minimized when taking the difference $dt$; this has been demonstrated also, e.g., by \cite{marinfranch2009}, who found that the precision in the estimate of differential ages in globular clusters can be pushed to 2-7\% including systematic errors. From this point of view, two methods that give slightly different absolute ages (depending on different assumptions) but compatible differential ages, will be completely equivalent for what concerns CCs; this has been well demonstrated in \cite{jiao2023}, where age measurements from FSF and Lick indices obtained on the same sample were compared, and even if there was some small offset in absolute ages (accountable to different treatments of the SFH), the relative ages were totally compatible, yielding $H(z)$ measurements in perfect agreement.
\end{remark}

\begin{remark}
One of the strengths of this method, as presented in Sect.~\ref{chap1:sec2}, is to rely on minimal premises and to provide measurements of the expansion history of the Universe independent of cosmological assumptions. For this reason, however, it is crucial to rely on a {\bf cosmology-independent estimate of the age}. While this may sound obvious (otherwise one would simply retrieve the assumed fiducial cosmology in a circular argument), in most astrophysical analyses it is not the case. Actually, to allow a better convergence of the results and to break some degeneracies, in most of the codes that measure galaxy ages a cosmological prior is considered, so that the maximum age of a galaxy is forced to be lower than the age of the Universe at that redshift (assuming a cosmology). For this reason, to study CC as cosmological probes it is crucial to derive an age estimate without any cosmological prior; this means that available codes should be adapted and revised to implement this feature \citep[as done, e.g., in][]{jiao2023, tomasetti2023}.
\end{remark}

\begin{remark}
In the CC approach, the basic quantity to be constrained is the slope of the age--redshift relation, $dt/dz$. This means that any potential systematic effect that could alter this relation could induce, as a consequence, a bias in the measurement of $H(z)$; these should, therefore, be minimized. Several possible systematic effects might produce this effect, typically involving an inclusion in the sample of a population younger than the average. Among those, we mention in particular the {\bf progenitor bias} and {\bf rejuvenation}. The progenitor bias \citep{franx1996,vandokkum2000} occurs when, given a fixed selection criterion, younger objects than high-redshift ones are progressively included in the sample with decreasing redshift (since young galaxies formed later than high-redshift ones might eventually enter the selection). Rejuvenation happens, instead, when a passive population is contaminated by a younger one due to some activity that triggers a new event of star formation. Both these biases are actually minimized by the differential approach (since the quantity $dt$ has to be estimated between two CC samples close in redshift, minimizing the progenitor bias) and by the selection procedure described in Sect.~\ref{chap1:sec3} (because the adopted combined criteria minimize the potential contamination by even a small component of star-forming sub-population, and also guarantees a tight homogeneity of the sample in terms of formation time).
\end{remark}

\begin{remark}
All the age estimates rely in some way on some {\bf Stellar Population Synthesis (SPS) models}, which allows us to interpret some observational features in terms of physical parameters (such as metallicity and age). For this reason, it is crucial to consider this in the total error budget of the analysis, as will be discussed in Sect.~\ref{chap1:sec5}.
\end{remark}
\end{BoxTypeA}

\section{From relative ages to the expansion rate of the Universe}\label{chap1:sec5}

In Sect.~\ref{chap1:sec4} it has been presented how to derive relative ages $dt$ for a sample of cosmic chronometers. The final step needed is to convert relative ages to a measurement of the Universe's expansion rate. As introduced above, the CC method relies on a measurement of the differential age evolution in a redshift interval, but we still have to discuss what are the best ways to define these intervals to obtain the optimal cosmological measurements.

\begin{BoxTypeA}{Binning the data}

\noindent There are different ways and approaches to derive $dz/dt$ from the data. Here below are summarized three different approaches with which it is possible to bin the data to derive the differential age in a redshift interval.\\~\\

\begin{itemize}
\item {\bf Stacking or single spectra analysis?} Depending on the considered method to estimate the age and on the quality of data, it could be more convenient to work on single objects or to average galaxies in specific bins. When spectra have a good SNR and resolution \citep[as, e.g., in][]{borghi2022, jiao2023, tomasetti2023}, the optimal strategy is to perform the analysis on an object-by-object basis: this enables much better granularity in the analysis and does not smooth out the potentially slightly different properties of the selected CC. On the other hand, if the data available have lower quality, it is still possible to apply the CC method, in this case increasing the SNR by averaging objects together. A common approach is to stack together the spectra so that the SNR can be increased by a factor $\sqrt{N}$; in this case, the only point of attention is to choose bins of galaxy properties so that the objects that are averaged together are not too dissimilar, not to increase the variance in the final product (see also the next points). Moreover, stacked spectra can also be studied to check the performances of single-galaxy analysis \citep[see, e.g.,][]{moresco2012, moresco2016}. 
\item {\bf Choice of the redshift interval.} Once decided to perform the analysis on single or stacked spectra, it is possible to construct the age--redshift (or $D4000$-redshift) relation. At this point, it is necessary to bin the data in redshift intervals, average the measured age in each interval, and then use it to derive the slope (which will provide $dz/dt$). The choice of the way to best divide the data in redshift intervals will depend on the data available, and it is always a trade-off between two contending effects. On the one side, the more redshift bins, the more $H(z)$ points it will be possible to derive, hence reconstructing the expansion rate of the Universe in more detail; on the other side, the more redshift bins, the fewer points will be averaged in each bin, hence the scatter will be larger and the accuracy in the determination of $dz/dt$ smaller. A good rule of thumb is to choose the redshift interval such that the evolution in age is larger than the scatter in the measurement, to be able to robustly determine the slope of the averaged age--redshift relation \citep[more discussion on this topic can be found, e.g., in][]{moresco2012, moresco2016, borghi2022, tomasetti2023}.
\item {\bf Analysis in different mass/velocity dispersion bins.} As discussed in Sect.~\ref{chap1:sec3}, a mass/velocity dispersion cut is needed for an optimal CC selection. However, as highlighted by \cite{thomas2010}, even with a high-mass threshold selection, galaxies with different masses will have on average different redshift of formation according to the mass-downsizing scenario. For this reason, to ensure the maximum homogeneity in the formation time of the targets, galaxies are typically divided into mass or velocity dispersion bins before performing the analysis. In this way, one will end up with subsamples with slightly different absolute ages (as shown in Fig.~\ref{fig:CCmethods}) but an extremely coherent slope, ensuring an unbiased estimate of $H(z)$ that can, potentially, be averaged a-posteriori in each redshift bin \citep[as done, e.g., in][]{moresco2012,moresco2016, borghi2022, jiao2023, tomasetti2023}.
\end{itemize}

\end{BoxTypeA}

Current CC measurements of the Universe's expansion history are presented in Tab.~\ref{chap1:tab1}, and shown in the right column of plots in Fig.~\ref{fig:CCmethods}. These data comprise values obtained spanning the wide range of methods described in Sect.~\ref{chap1:sec4}. After the seminal work by \cite{jimenez2002}, the first application of the CC method on real data was done by \cite{simon2004}; analyzing a combination of passive galaxies selected from the SDSS early data release and GDDS surveys with an FSF approach, they derived 8 $H(z)$ measurements in the range $0<z<1.75$. Later on, also \cite{zhang2014}, \cite{ratsimbazafy2017}, \cite{jiao2023}, and \cite{tomasetti2023}, applying the same technique on different surveys (SDSS Data Release Seven, 2dF–SDSS, LEGA-C, and VANDELS, respectively) obtained 7 additional measurements of the Hubble parameter from $z\sim 0.07$ to $z\sim1.26$.
The method based on the $D4000$ was instead introduced by \cite{moresco2011}, and adopted in \cite{moresco2012}, \cite{moresco2015}, and \cite{moresco2016}. In the three papers, several surveys were exploited, namely the SDSS Data Release 6 Main Galaxy Sample, the SDSS Data Release 7 Luminous Red Galaxy sample, zCOSMOS, K20, UDS, and SDSS BOSS Data Release 9. In these analyses, more than $10^5$ CCs were selected, providing 15 new measurements at $0.18<z<2$. Taking advantage of the exquisite resolution and SNR of the LEGA-C survey, \cite{borghi2022} exploited for the first time the possibility of applying the CC method based on age measurements from Lick indices, deriving a new constraint on $H(z)$ at $z\sim0.7$. The accurate results obtained in this analysis have also been used to explore a new path to determine the Hubble parameter through a Machine-Learning approach in \cite{jimenez2023}, training a neural network to infer stellar ages from the wide photometric coverage (from the UV to the near-IR rest-frame wavelengths) of the COSMOS2015 survey; in this way, a new $H(z)$ value has been determined at $z\sim0.75$.

\begin{table}[t]
\TBL{\caption{Measurement of the Hubble expansion rate obtained with CC, in units of \Hunit. For each measurement, we report the method used to determine the relative age (see Sect.~\ref{chap1:sec4}) with the following symbols: full spectral fitting (F), D4000 (D), Lick indices (L), Machine Learning (ML).}\label{chap1:tab1}. }
{\begin{tabular*}{\textwidth}{@{\extracolsep{\fill}}@{}lllll|lllll@{}}
\toprule
\multicolumn{1}{@{}l}{\TCH{$z$}} &
\multicolumn{1}{l}{\TCH{$H(z)$}} &
\multicolumn{1}{l}{\TCH{$\sigma_{H(z)}$}} &
\multicolumn{1}{l}{\TCH{method}} &
\multicolumn{1}{l}{\TCH{reference}} &
\multicolumn{1}{l}{\TCH{$z$}} &
\multicolumn{1}{l}{\TCH{$H(z)$}} &
\multicolumn{1}{l}{\TCH{$\sigma_{H(z)}$}} &
\multicolumn{1}{l}{\TCH{method}} &
\multicolumn{1}{l}{\TCH{reference}}\\
\colrule
0.07 & 69.0 & 19.6 & F & \cite{zhang2014} & 0.593 & 104 & 13 & D & \cite{moresco2012}\\
0.09 & 69 & 12 & F & \cite{simon2004} & 0.68 & 92 & 8 & D & \cite{moresco2012}\\
0.12 & 68.6 & 26.2 & F & \cite{zhang2014} & 0.75 & 98.8 & 33.6 & L & \cite{borghi2022}\footnotemark{a}\\
0.17 & 83 & 8 & F & \cite{simon2004} & 0.75 & 105 & 10.76 & ML & \cite{jimenez2023}\footnotemark{a}\\
0.179 & 75 & 4 & D & \cite{moresco2012} & 0.781 & 105 & 12 & D & \cite{moresco2012}\\
0.199 & 75 & 5 & D & \cite{moresco2012} & 0.8 & 113.1 & 25.22 & F & \cite{jiao2023}\footnotemark{a}\\
0.20 & 72.9 & 29.6 & F & \cite{zhang2014} & 0.875 & 125 & 17 & D & \cite{moresco2012}\\
0.27 & 77 & 14 & F & \cite{simon2004} & 0.88 & 90 & 40 & F & \cite{stern2010}\\
0.28 & 88.8 & 36.6 & F & \cite{zhang2014} & 0.9 &  117 &  23 & F & \cite{simon2004}\\
0.352 & 83 & 14 & D & \cite{moresco2012} & 1.037 & 154 & 20 & D & \cite{moresco2012}\\
0.38 & 83 & 13.5 & D & \cite{moresco2016} & 1.26 & 135 & 65 & F & \cite{tomasetti2023}\\
0.4 & 95 & 17 & F & \cite{simon2004} & 1.3 & 168 & 17 & F & \cite{simon2004}\\
0.4004 & 77 & 10.2 & D & \cite{moresco2016} & 1.363 & 160 & 33.6 & D & \cite{moresco2015}\\
0.425 & 87.1 & 11.2 & D & \cite{moresco2016} & 1.43 & 177 & 18 & F & \cite{simon2004}\\
0.445 & 92.8 & 12.9 & D & \cite{moresco2016} & 1.53 & 140 & 14 & F & \cite{simon2004}\\
0.47 & 89.0 & 49.6 & F & \cite{ratsimbazafy2017} & 1.75 & 202 & 40 & F & \cite{simon2004}\\
0.4783 & 80.9 & 9 & D & \cite{moresco2016} & 1.965 & 186.5 & 50.4 & D & \cite{moresco2015}\\
0.48 & 97 & 62 & F & \cite{stern2010} & & & & & \\
\botrule
\end{tabular*}}{%
\begin{tablenotes}
\footnotetext[a]{measurements with different methods but on the same (or calibrated on the same) sample, avoid using these data jointly in an analysis}
\end{tablenotes}
}%
\end{table}

\subsection{Systematic uncertainties}
\label{chap1:sec5:sub1}

Together with statistical uncertainties (connected to the quality of the data and, eventually, to the binning procedure chosen), several other effects may contribute to the total error budget of the CC technique; they are linked to the various stages of the method described previously, and in particular to:
\begin{itemize}
\item some residual contamination of young, star-forming outliers in the selection procedure, biasing the slope of the age--redshift relation;
\item a non-perfectly broken age--metallicity degeneracy, so  that a prior on the metallicity of the population (with associated uncertainty)  is needed to be assumed to apply the method;
\item the dependence on the assumed SPS model to derive information on the stellar age, which, in turn, can depend on the particular ingredients of the model such as the Initial Mass Function (IMF) and the stellar library considered. Currently, most measurements have been performed considering SPS models by \cite{bruzual2003} or \cite{maraston2011}, but in \cite{moresco2020} more models have been considered to estimate their impact on the analysis \citep{conroy2009,conroy2010,vazdekis2016};
\item any further effects that may potentially bias the determination of the slope of the age--redshift relation, such as the progenitor bias, rejuvenation, or dependence on the assumed mass cuts.\newline
\end{itemize}

Some of these issues, as discussed above in Sect.~\ref{chap1:sec4}, are minimized by the fact that the CC method is based on a differential measurement, and the quantity $dz/dt$ is estimated in redshift intervals so close that the intrinsic evolution of the properties of the sample is negligible with respect to the age evolution of the Universe. On the other hand, systematic errors have to be carefully taken into account to obtain reliable estimates of $H(z)$. \cite{moresco2020} explored in detail all these possible contributions, and explicitly provided a recipe to obtain the total covariance matrix for CC analysis as:
\begin{align} 
{\rm Cov}_{ij}^{\rm tot} = & {\rm \;\;Cov}_{ij}^{\rm stat} 
 & {\rm \it statistical\;errors}\nonumber\\ 
& + {\rm Cov}_{ij}^{\rm met} + {\rm Cov}_{ij}^{\rm young}& {\rm \it systematic\;errors\;(metallicity\;and\;contamination)}\nonumber\\
& + {\rm Cov}_{ij}^{\rm SFH}+{\rm Cov}_{ij}^{\rm IMF}+{\rm Cov}_{ij}^{\rm st. lib.}+{\rm Cov}_{ij}^{\rm SPS}& {\rm \it systematic\;errors\;(SPS\;model)}
\label{eq:totcovCC}
\end{align}
where the different components of the error budget have been split into various rows: the first one shows the statistical part, the second one the systematic due to the uncertainty in metallicity and contamination by a young component (rejuvenation), and the third one the systematic related to SPS modeling.
Each of these contributions needs to be, at first impact, minimized through an as accurate as possible CC selection and by choosing the proper bins, but then properly estimated depending on the procedure adopted in each analysis. In Tab.~\ref{chap1:tab1} are reported only the statistical uncertainties, and the systematic ones have to be added to the total error budget as shown in Eq.~\ref{eq:totcovCC} before any cosmological analysis\footnote{In \url{https://gitlab.com/mmoresco/CC_covariance} are provided jupyter notebooks that illustrate how to properly estimate the CC total covariance matrix and include it in a cosmological analysis.}.
At the moment, the error budget is dominated by systematic errors due to metallicity uncertainty and SPS models. In general, with CC it is possible to reach an accuracy on the determination of the expansion history of the Universe around 5\% at $z\sim0.5$ up to 10-20\% at $z\sim2$.

\subsection{Cosmological constraints}
\label{chap1:sec5:sub2}

\begin{figure}[t]
\centering
\includegraphics[width=.95\textwidth]{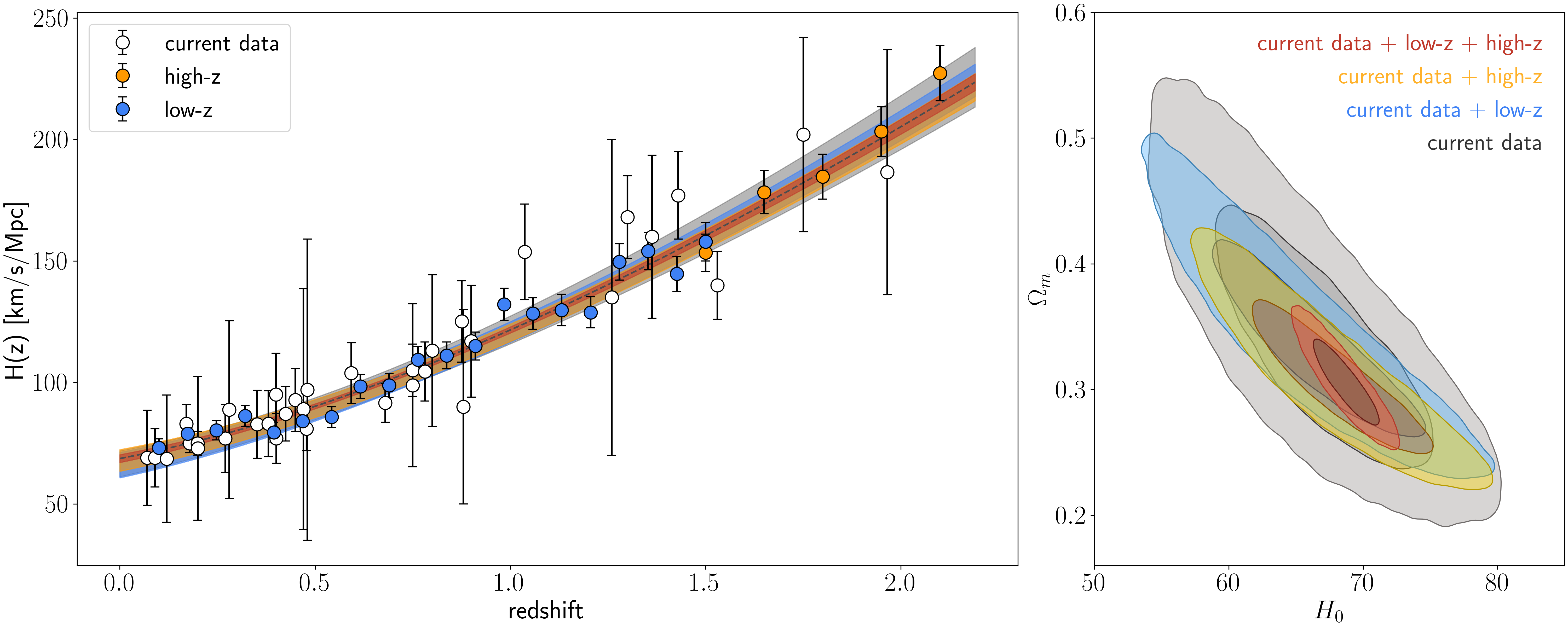}
\caption{Cosmological constraints from CC. In the left panel is shown, in white points, the current CC dataset as reported in Tab.~\ref{chap1:tab1}. Colored points are, instead, forecasts for future CC analyses \citep[following ][]{moresco2022} from low-z and high-z samples, as discussed in the text. The shaded regions show the 68\% confidence interval obtained by fitting with a flat $\Lambda$CDM model current data (in grey), current data combined with low-z (in blue) or high-z samples (in yellow), and the combination of all datasets. The constraints in the $H_0$--$\Omega_m$ plane can be found in the right panel, with the same color scheme.}
\label{fig:cosmoconstraints}
\end{figure}

Providing cosmology-independent estimates of the Hubble parameter $H(z)$, cosmic chronometers are powerful tools for obtaining information complementary to standard probes on how the Universe is expanding and, as a consequence, on its geometry and contents. While not directly constraining the Hubble constant (which is the value of the Hubble parameter at $z=0$), CC data could also be exploited to derive it indirectly, through extrapolation or other techniques discussed below, and hence give crucial additional pieces of evidence that can help in disentangling and understanding the current cosmological tensions between late-Universe and early-Universe probes \citep{verde2019,divalentino2021}.

One of the main strengths of CCs as cosmological probes is that they directly probe the Hubble parameter rather than some combination of its integral, like luminosity or angular diameter distances as for SNe or BAO; from this point of view, the advantage of this approach is that it has a higher sensitivity to cosmological parameters \citep[as reported by][]{jimenez2002}. Cosmological constraints can be derived from CC data following two general approaches. The first possibility is to fit the data with some assumed cosmological model and determine its corresponding parameters \citep{moresco2011,moresco2012b,moresco2016b}. Alternatively, it is possible to use model-independent approaches (like Gaussian Processes or Pad\'e approximation) to extrapolate the shape of the expansion of the Universe, allowing in this way an indirect measurement of the Hubble constant \citep{seikel2012,protopapas2014,montiel2014,haridasu2018,gomezvalent2018,capozziello2019,sun2021,bonilla2021,colgain2021b}; the most recent constraint, considering the full covariance matrix, on the Hubble constant obtained with this approach found $H_0=70.7\pm6.7$ \Hunit~\citep{favale2024}, and a similar result can be obtained directly fitting a $\Lambda$CDM model \citep{moresco2022}. Other proposed approaches involve the use of different estimators or Machine-learning-based algorithms \citep{arjona2020,mehrabi2021}. 

An application of CC data for a cosmological analysis is shown in Fig.~\ref{fig:cosmoconstraints}. The full compilation reported in Tab.~\ref{chap1:tab1} is considered (as shown in the left plot), as well as forecasts of $H(z)$ with CC from future surveys \citep[following][]{moresco2022}. Two scenarios are simulated: a ``low-z'' sample at redshifts $0<z<1.5$ that could be provided, e.g., by the SDSS BOSS and eBOSS \citep{eBOSS2020} and the Dark Energy Spectroscopic Instrument \citep[DESI,][]{DESI2016a} surveys, to mention currently available ones, or, considering also future missions and projects under study, the Astrophysics Telescope for Large Area Spectroscopy Probe \citep[ATLAS Probe,][]{Atlasmission} or the Wide-field Spectroscopic Telescope \citep[WST,][]{WST2024}; and a ``high-z'' sample at redshifts $1.5<z<2.1$ that will be provided by the ESA survey Euclid \citep{Euclid}. Leveraging on the large sample statistics that can be built upon these surveys, we consider here to be capable of measuring 20 $H(z)$ points for the low-z sample \citep[see][]{WST2024} and 5 for the high-z sample, with a statistical error of 5\% (compatible with the one reachable today). In this analysis, as indicated by \cite{moresco2020} we also include in the total error budget the contribution of systematic errors for the IMF and SPS models. In the right panel of Fig.~\ref{fig:cosmoconstraints} are shown the constraints that can be obtained by fitting in a flat $\Lambda$CDM model (where $H_0$ and $\Omega_m$ are free) the current dataset, alone or combined with either the two simulated samples and a full combination of all data where, for illustrative purposes, a more optimistic systematic error is considered\footnote{Here, as done in \cite{moresco2022}, it is assumed that the systematic uncertainty due to the different SPS models could be significantly reduced with future spectroscopic analyses and observation, that could help in identifying and reducing the number of SPS models that best reproduce the data.}.

As noticed above, current CC data do not allow us to weigh in significantly in the Hubble tension debate, providing an error on $H_0$ of the order of $\sim$8--9\% and of $\sim$20\% on $\Omega_m$. Interestingly, future data could potentially significantly change the situation, allowing us to map in much better detail the evolution of the Hubble expansion, at least up to $z\sim2-2.2$; in particular, with the low-z sample the accuracy on the Hubble constant could increase to 6--7\%, and with the full combination, considering the improvement in CC modeling, could reduce down to $\sim$2\% (and to $\sim$7\% on $\Omega_m$). In this case, CC alone will provide fundamental and valuable information that can constitute an independent piece of the puzzle to reconstruct and disentangle cosmological mysteries.

In parallel to that, several works proved also the strength of CC in combination with other cosmological probes, due to their different sensitivity and degeneracies between parameters. Several works pursued this possibility, exploring different combinations between probes, demonstrating that: {\it i)} when combined with SNe the statistical power of CC is similar to the one of BAO, presenting a similar degeneracy between parameters \citep[e.g.,][]{moresco2012b,moresco2016b}, {\it ii)} when combined to other probes, CCs provide a pivot value on the Hubble parameter that cannot be given by other probes (e.g. BAO, SNe, or GRB) and help in constraining cosmological parameters \citep{zhao2017, lin2020,lin2021,cogato2024}, and {\it iii)} providing a cosmology-independent constraint CCs are in particular useful to test and study all kind of cosmological models \citep[see e.g.][]{bilicki2012,nunes2016,capozziello2018,colgain2019, capozziello2019b,vonmarttens2019,yang2019,benetti2019,aljaf2021,ayuso2021,reyes2021,benetti2021,vagnozzi2021,gonzalez2021}. In this case, the accuracy of the determination of the Hubble constant is increased; taking advantage of the synergy between all the various probes, and their latest measurements (including CC, SNe, BAO, and Gamma-Ray Burst) gives $H_0=66.9^{+3.5}_{-3.4}$ \Hunit~ \citep{cogato2024}.

\section{Conclusions and future prospects}\label{chap1:sec6}

Cosmic chronometers represent a novel, powerful cosmological probe that can provide cosmology-independent constraints on the expansion history of the Universe. The great advantages of this method are that it gives a direct measurement of the Hubble parameter without the need for any extrapolation or cosmological assumption, and that, being based on a differential approach, it only needs accuracy and robustness in the determination of the differential ages $dt$ to be applied. At the moment, the main limitations of the method are the fact that there does not exist a dedicated survey to detect CCs (as, for example, for CMB, SNe, or BAO) and that current errors are dominated by systematics driven by the uncertainties in the adopted SPS models, as discussed in Sect.~\ref{chap1:sec5:sub1}. 

The consequence of the first issue is the fact that current measurements have been obtained as legacy or serendipitous data from surveys designed for other purposes, never being optimized for this science case. In the future, however, several spectroscopic surveys are expected to increase by orders of magnitude the number of passive galaxies detected at high resolution and SNR, like ATLAS Probe \citep{Atlasmission} and WST \citep{WST2024}, and many others are already taking data, like Euclid \citep{Euclid}, SDSS \citep{eBOSS2020}, and DESI \citep{DESI2016a}. This will provide huge statistics that will allow us for the first time to derive extremely accurate constraints up to $z\sim2$ minimizing the statistical errors.

On the other hand, to make progress it is also crucial to make a step forward in the modelization, in particular by reducing all possible sources of systematic errors. While, on the one side, the quality of future data will allow us a continuously better and more accurate selection of a pure sample of CC, on the other side, current instruments are starting to provide observations for massive and passive galaxies with excellent quality in terms of spectral resolution, SNR, wavelength and redshift coverage; as a few examples, we may think of the data provided by the James Webb Space Telescope \citep[JWST,][]{JWST}, or from the spectrographs X-Shooter and Moons. These instruments, as discussed in \cite{moresco2020}, will provide us data that not only will enable us to do very accurate determinations of $dt$, but will also be the perfect test-bed for checking and comparing models, reducing our theoretical uncertainties and, as a consequence, the systematic errors. 
For these reasons, the future of cosmic chronometers is bright, and they could be powerful independent cosmological probes to shed light on the dark nature of our Universe.


\begin{BoxTypeA}{Key points}
\begin{itemize}
\item Cosmic chronometers are a novel cosmological probe that provides a measurement of the expansion history of the Universe, the Hubble parameter $H(z)$, without relying on any cosmological assumption, by only measuring the differential age evolution $dt$ of tracers of the age evolution of the Universe:
\begin{equation}
    H(z)=-\frac{1}{(1+z)}\frac{dz}{dt}\;\; .\nonumber
\end{equation}
\item The best cosmic chronometers have been identified as very massive ($\log_{10}(M/M_{\odot})>$10.5--11) and passively evolving galaxies because they represent the oldest, most homogeneous, and synchronous population found at each redshift.
\item To apply the method, it is crucial to ensure the purity of the CC sample by combining different selection criteria (photometric, spectroscopic, stellar mass cut), minimizing the contamination by young, star-forming outliers.
\item The method is based on a {\it differential approach} since the important quantity to be measured is the relative and not the absolute age. For this reason, provided that the measurement of the relative ages is robust, differences in the absolute ages factor out and do not impact the method. It is of extreme importance, though, that the ages are estimated without assuming any cosmological prior.
\item While the differential approach and selection process help to minimize several sources of uncertainty, it is of extreme importance to consider in the analysis the total covariance matrix as provided by Eq.~\ref{eq:totcovCC}, which accounts for all possible contributions to the statistical and systematic errors.
\item This method provides a determination of the Hubble parameter with an accuracy of around 5\% at $z\sim0.5$ up to 10-20\% at $z\sim2$. The current constraint on the Hubble constant based on CC data alone reaches an accuracy of $\sim$8--9\% ($\sim$3\% when combined with other cosmological probes), that could be reduced to $\sim$2\% with future data and improvements in the treatment of systematics.
\end{itemize}
\end{BoxTypeA}


\begin{ack}[Acknowledgments]

The author acknowledges support from MIUR, PRIN 2017 (grant 20179ZF5KS) and PRIN 2022 (grant 2022NY2ZRS\_001), and the grants ASI n.I/023/12/0 and ASI n.2018-23-HH.0.  \end{ack}

\bibliographystyle{Harvard}
\bibliography{reference}

\end{document}